\documentclass[preprint2]{aastex62}
\usepackage{bm}

\renewcommand{\aa}{\bm{a} }
\newcommand{\bb}{\bm{b} }

\graphicspath{{./}{figures/}}

\received{XXX}
\revised{XXX}
\accepted{XXX}
\submitjournal{ApJ Letters}

\shorttitle{Shock-Transmitted Turbulent Structures}
\shortauthors{Trotta et al.}

\begin{document}

\title{On the transmission of turbulent structures across the Earth's Bow Shock}

\correspondingauthor{Domenico Trotta}
\email{d.trotta@imperial.ac.uk}

\author[0000-0002-0608-8897]{Domenico Trotta}
\affil{The Blackett Laboratory
Imperial College London
London. SW7 2AZ, UK}

\author[0000-0003-4168-590X]{Francesco Pecora}
\affil{Department of Physics and Astronomy, University of Delaware,
Newark DE 19716 USA}

\author[0000-0003-1821-7390]{Adriana Settino}
\affil{Dipartimento di Fisica, Universit\`a della Calabria, I-87036 Cosenza, Italy}

\author[0000-0003-1059-4853]{Denise Perrone}
\affil{ASI - Italian Space Agency, Via del Politecnico snc, I-00133 Rome, Italy
}

\author[0000-0002-3039-1255]{Heli Hietala}
\affil{The Blackett Laboratory
Imperial College London
London. SW7 2AZ, UK}

\author[0000-0002-7572-4690]{Timothy Horbury}
\affil{The Blackett Laboratory
Imperial College London
London. SW7 2AZ, UK}

\author[0000-0001-7224-6024]{William Matthaeus}
\affil{Department of Physics and Astronomy, University of Delaware,
Newark DE 19716 USA}

\author[0000-0002-8175-9056]{David Burgess}
\affil{School of Physics and Astronomy
Queen Mary University of London,
London, E1 4NS, UK}

\author[0000-0001-8184-2151]{Sergio Servidio}
\affil{Dipartimento di Fisica, Universit\`a della Calabria, I-87036 Cosenza, Italy}

\author[0000-0002-1296-1971]{Francesco Valentini}
\affil{Dipartimento di Fisica, Universit\`a della Calabria, I-87036 Cosenza, Italy}



\begin{abstract}

Collisionless shocks and plasma turbulence are crucial ingredients for a broad range of astrophysical systems. The shock-turbulence interaction, and in particular the transmission of fully developed turbulence across the quasi-perpendicular Earth's bow shock, is here addressed using a combination of spacecraft observations and local numerical simulations. An alignment between the Wind (upstream) and MMS (downstream) spacecraft is used to study the transmission of turbulent structures across the shock, revealing an increase of their magnetic helicity content in its downstream. Local kinetic simulations, in which the dynamics of turbulent structures is followed through their transmission across a perpendicular shock, confirm this scenario, revealing that the observed magnetic helicity increase is associated with the compression of turbulent structures at the shock front.

\end{abstract}



\section{Introduction}
\label{sec:intro}


The interaction between collisionless shocks and plasma turbulence is a key ingredient for the understanding of many astrophysical phenomena \citep[e.g.,][]{Bykov2019, Guo2021}. Such an interaction has been of growing interest in recent literature, involving theoretical \citep[][]{Zank2002}, numerical \citep[][]{Giacalone1996,Guo2012,Trotta2021} and observational \citep[][]{Lario2019,Pitna2017,Perri2021} efforts.

A common example of such an interaction is the Earth's bow shock, at the interface between the turbulent solar wind flow and the Earth's magnetosphere. From the first \emph{Pioneer} spacecraft observations \citep{Dungey1979} to the modern,  Magnetospheric MultiScale (MMS) mission \citep{Burch2016}, an extraordinary amount of knowledge has been built around the Earth's bow shock, supported by numerical \citep[e.g.,][]{Turc2018} and theoretical modelling \citep[e.g.][]{Chapman2003}. Earth's bow shock has become a prototype to address energy conversion and particle acceleration \citep[see ][]{Schwartz2021}.

An exceptionally important property of turbulent flows is the emergence
of coherent structures, that characterize the energy transfer from large to small scales \citep[e.g.][]{Holmes1996,GrecoEA10,ServidioEA15,Krieger2020}. Such structures, ubiquitous in the heliosphere, are fundamental building blocks of the interplanetary environment \citep[see][for a review]{Khabarova2021,Pezzi2021}. Particular attention has recently been dedicated to the study of such turbulent structures' magnetic helicity content, an important rugged invariant of magnetohydrodynamics (MHD) theory, providing important information about the topology of the magnetic field \citep[][]{matthaeus1982evaluation, pecora2021identification}.

The transmission of turbulent structures across collisionless shocks is an important factor in shock energetics \citep[e.g.,][]{Zank2021}, as turbulent structures themselves are important particle accelerators \citep[e.g.][]{Drake2006, LeRoux2018, Trotta2020b}, and their dynamics are affected by shock crossing. On the other hand, transmission of turbulent structures may induce strong changes on the shock transition properties \citep[e.g.][]{Kajdic2019,Trotta2021}, as well as affecting particle transport across the shock transition \citep[e.g.,][]{Perri2015}.

In the present work, we tackle the problem of the transmission of turbulent structures across quasi-perpendicular shocks. We use a combination of Wind and MMS data to study the turbulent properties of the plasma upstream and downstream of the Earth's bow shock. We then present novel kinetic simulations of shocks propagating in turbulent media, introducing the possibility to follow turbulent structures through their interaction with the shock. We find strong helicity increase from the shock upstream to the downstream, and propose a simple theoretical treatment that links the time evolution of magnetic helicity to plasma compression in the MHD framework. The paper is organized as follows: in Section \ref{sec:obsrevations} the spacecraft observations are presented, and the structures' transmission across the bow shock is discussed in Section \ref{sec:transmission}. In Section \ref{sec:simulations} the simulations results are shown. Finally, the conclusions are reported in Section \ref{sec:conclusions}.

\begin{deluxetable*}{ccccccccc}
\tabletypesize{\footnotesize}
\tablecolumns{9}
\tablewidth{0pt}
\tablehead{
\colhead{Mission} & \colhead{start (UT)} & \colhead{end (UT)} & \colhead{B (nT)} & \colhead{N$_i$ (cm$^{-3}$)} & \colhead{V$_i$ (km/s)}& \colhead{T$_i$ (eV)} & \colhead{$\beta_i$} & \colhead{v$_A$ (km/s)}}
\startdata
Wind - Solar wind & 20:30:16 & 01:59:53 & 10 $\pm$ 1 & 12 $\pm$ 3 &  390 $\pm$ 17 & 12 $\pm$ 2 & 0.7 $\pm$ 0.3 & 63 $\pm$ 12\\
MMS1 - Upstream & 22:08:50 & 22:09:10 & 11 $\pm$ 2 & 8.9 $\pm$ 0.7 & 345 $\pm$ 6 & 47 $\pm$ 22 & 1.3 $\pm$ 0.6 & 83 $\pm$ 3\\
MMS1 - Downstream & 22:09:50 & 22:10:10 & 32 $\pm$ 4 & 37 $\pm$ 5 & 186 $\pm$ 15 & 118 $\pm$ 31 & 1.8 $\pm$ 0.8 & 114 $\pm$ 3\\
\enddata
\caption{Summary of intervals used for Wind and MMS1 (rows) \emph{in-situ} analyses on December 31, 2017: Start and end times, magnetic field, ion density, bulk flow speed, temperature, beta and Alfv\'en speed.\label{tab:UP-DW}}
\end{deluxetable*}

\section{Spacecraft observations}
\label{sec:obsrevations}
In this section, we compare the statistical properties of turbulence in the pristine solar wind with the shocked magnetosheath plasma. These analyses, which are not intended to highlight a 1 to 1 correspondence between the solar wind and the magnetosheath structure transmission, reveal important statistical differences that probably arise due to the interaction between the shock-front and the incoming solar wind turbulence. In this regard, the structure/shock interaction might play a crucial role.

\subsection{Wind and MMS data}
\label{subsec:data}
On 2017 December 31, during a crossing from the pristine solar wind to the magnetosheath plasma, MMS observed a quasi-perpendicular bow-shock on the dusk side. Properties and geometry of the bow shock have been evaluated by averaging MMS1 measurements in the upstream and downstream regions using the same time interval ($20$~s) and carefully excluding the shock and foot regions (see Table~\ref{tab:UP-DW}).

We used magnetic field data from the FluxGate Magnetometer (FGM) \citep[][]{russell2016}, sampled at $16$~Hz in fast mode and the particle data collected by the Fast Plasma Investigation (FPI) instrument \citep{Pollock2016} with a cadence of $4.5$~s in fast mode. In the downstream region of Earth's bow shock, MMS observed a continuous interval (about $4$~hours) of shocked solar wind. In the meanwhile, both DSCOVR and Wind were located on the same side of MMS, but hundreds of R$_E$ upstream of Earth's bow shock. During this time, both DSCOVR and Wind measurements are consistent, and we decide to use Wind data as a proxy for upstream conditions.
\begin{figure*}[ht!]
\centering
\includegraphics[width=\textwidth]{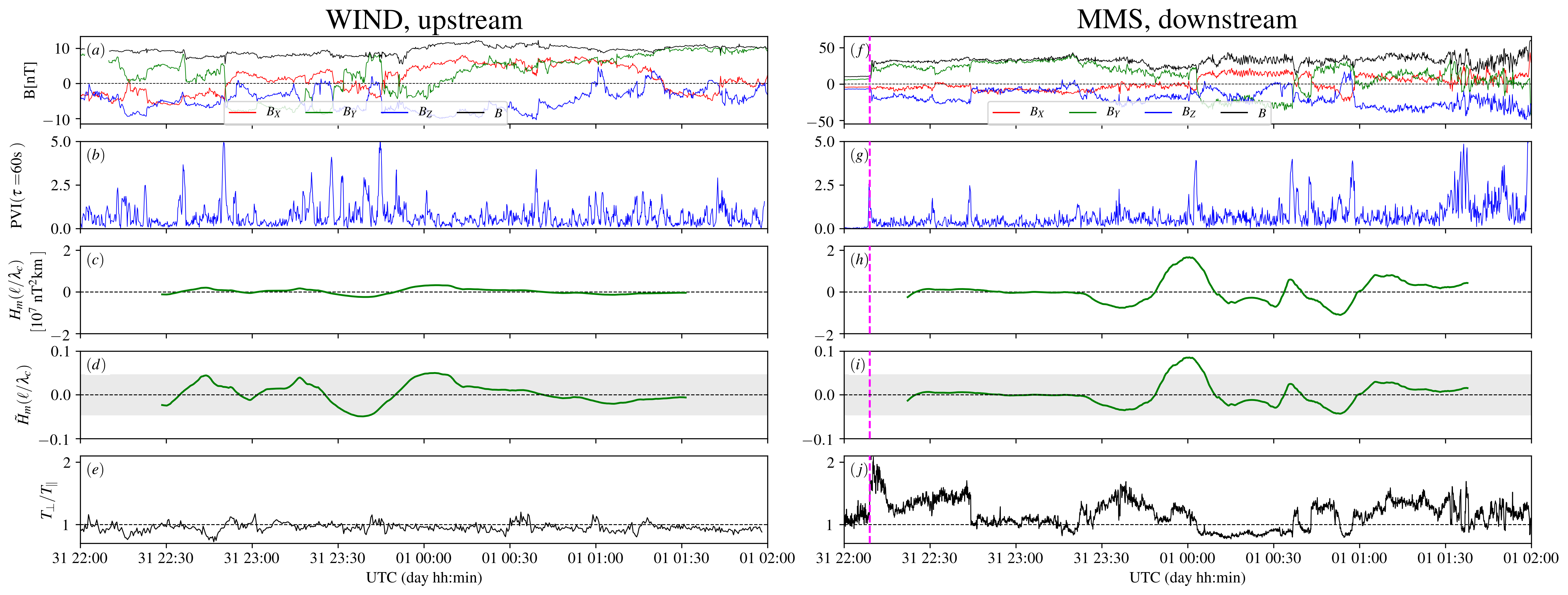}
\caption{Selected stream of solar wind passing through the Earth's bow shock for Wind (upstream, left) and MMS (downstream, right) satellites, showing magnetic field, PVI, magnetic helicity (normalised and unnormalised) and temperature anisotropy. The dashed magenta vertical line indicates the position of the bow shock.
\label{fig:BPHT}
}
\end{figure*}

As we considered a long time interval during which the bulk flow remains almost constant, we can reasonably assume that  that the key plasma parameters are similar at Wind and upstream of the shock at MMS. To further endorse this point, we have compared the mean values of the main quantities in the upstream intervals observed by Wind, which is upstream of MMS, with a timeshift of about 55 minutes. We observe a very good agreement between the two spacecraft measurements as shown in Table~\ref{tab:UP-DW}, thus suggesting that our previous assumption is reasonable, at least, in a statistical sense.

\subsection{Magnetic Helicity}
\label{sec:helicity}
Magnetic helicity is a rugged invariant of MHD turbulence, defined as $H_m = \langle \aa \cdot \bb \rangle$, where $\aa$ is the vector potential associated with magnetic field fluctuations $\bb={\bm \nabla} \times {\bm a}$. The averaging operation $\langle \dots \rangle$ is performed over an appropriate volume \citep{woltjer1958theorem, matthaeus1982measurement}.

Magnetic helicity can be used as a proxy to detect filamentary structures (such as flux tubes), exploiting their helical nature \citep[e.g.,][]{telloni2012wavelet, zhao2020identification}. In this work, we use the prescription for calculating magnetic helicity using single-spacecraft measurements, presented in \citet{pecora2021identification}, which builds up on the one proposed in early works by \citet{matthaeus1982evaluation, matthaeus1982measurement}. This method calculates a local (both in configuration and increment spaces) value  of $H_m$ at a certain point $x$, retaining the contribution of structures up to a scale $\ell$, namely
\begin{equation}
    H_m(x, \ell) = \int_0^\ell dl_i \; h(l_i) \int_{x-\frac{w_0}{2}}^{x+\frac{w_0}{2}} d\xi \; \epsilon_{ijk}b_j(\xi)b_k(\xi+l_i),
    \label{eq:Hm}
\end{equation}
where the \textit{i-th} direction of the increments is along the relative spacecraft-solar wind motion, and the components $b_j$ and $b_k$ are those perpendicular to it; the interval $w_0$ is chosen to be proportional to the local correlation length and $h(l) = \frac{1}{2}\left[ 1 + \cos\left(\frac{2\pi l}{w_0}\right) \right]$ is the Hann window (or any other differentiable filter) used to smooth fluctuations to zero at the boundaries of the considered interval.

From the quantity in Eq.~\ref{eq:Hm}, it is possible to derive a version of the magnetic helicity normalized to the local energy content, namely
\begin{equation}
\tilde{H}_m = \frac{H_m}{\langle \delta b^2 \rangle \lambda_c},
\label{eq:sig}
\end{equation}
where $\langle \delta b^2 \rangle$ is the fluctuation energy computed from a local average taken on an interval of about a correlation length $\lambda_c$. This latter normalization enhances the contribution of smaller helical-structures.

\section{Turbulence transmission across the bow shock}
\label{sec:transmission}

Figure~\ref{fig:BPHT} shows a comparison between Wind (left) and MMS observations (right), representing a proxy for the upstream and downstream bow shock plasma, respectively. The top panels show magnetic field measurements at one-minute resolution, for both Wind and MMS. The Partial Variance of Increments \citep[PVI,][]{Greco2008} is computed using these magnetic field streams, in panels (b) and (g).

Panels (c) and (h) of Fig.~\ref{fig:BPHT} show that magnetic helicity greatly increases when the solar wind passes from upstream to downstream. This, as shown below in simulations, indicates that filamentary structures present upstream get compressed by the shock and increase their helicity content. The present scenario can also be noticed by looking at the normalized helicity in panels (d) and (i). In the upstream region, the $\tilde{H}_m$ signal lies almost entirely within the (grey) shaded region that represents the values of helicity that cannot be readily distinguished from those pertaining to random fluctuations rather than orderly flux tubes. However, the larger values going beyond the ``uncertainty band'' suggest the presence of flux tubes, which is confirmed in the downstream signal (panel (i)). Note that panels (c), (h) and (d), (i) are on the same scale respectively. This statistical evidence of enhanced magnetic flux tubes in the magnetosheath have been observed in previous works \citep[e.g.,][]{Stawarz2021}.

Bottom panels (e) and (j) show the temperature anisotropy $T_\perp/T_\parallel$ calculated with respect to the mean magnetic field over the whole intervals. Upstream of the shock, where the helical signature is weaker, there is lower temperature anisotropy. In the downstream region, there is an increase of $T_\perp/T_{||}$, in agreement with some other observations of the magnetosheath in quasi-perpendicular regions \citep[e.g.,][]{Dimmock2015}. Note that the downstream temperature anisotropy is patchy, with an increase in the perpendicular temperature where helical structures have been compressed and enhanced.

The plasma behaviour in the shock downstream is characterised by an injection of fluctuations around ion kinetic scales, likely a consequence of typical plasma instabilities for perpendicular shocks \citep[e.g,][]{Burgess2015}. This was confirmed  by the analysis of magnetic field spectra (not shown here), indicating that kinetic effects are fundamental for the overall system and cannot be neglected.

\label{sec:simulations}

Spacecraft observations give important insights about the statistical nature of shock-processed turbulent structures. Clearly, it is not possible to follow the dynamics of a particular structure across the shock using \emph{in-situ} observations. This important experiment can instead be carried out using numerical simulations.

We perform kinetic simulations of perturbed and unperturbed perpendicular, collisionless shocks with and without pre-existing plasma turbulence. Our setup, introduced in \citet{Trotta2021}, uses 2.5D MHD simulations to generate turbulent fields \citep[e.g.,][]{Perri2017}, that are then used as an initial condition for a hybrid kinetic simulation (where protons are treated kinetically, and electrons as a fluid) that studies the shock propagation in the turbulent upstream. We impose magnetic field fluctuations amplitude to be $\delta B / B_0 = 0.4$, where $\mathbf{B}_0 = B_0 \, \hat{e}_z$  is the mean magnetic field. We use the HYPSI code \citep[see ][]{Trotta2019,Trotta2020a} for the modelling of the (perturbed) shock evolution, initiated using the injection method \citep[][]{Quest1985}.  The shock propagates in the negative $x$-direction, with the mean upstream flow along the shock normal.

\begin{figure}[ht!]
\centering
\includegraphics[width=0.46\textwidth]{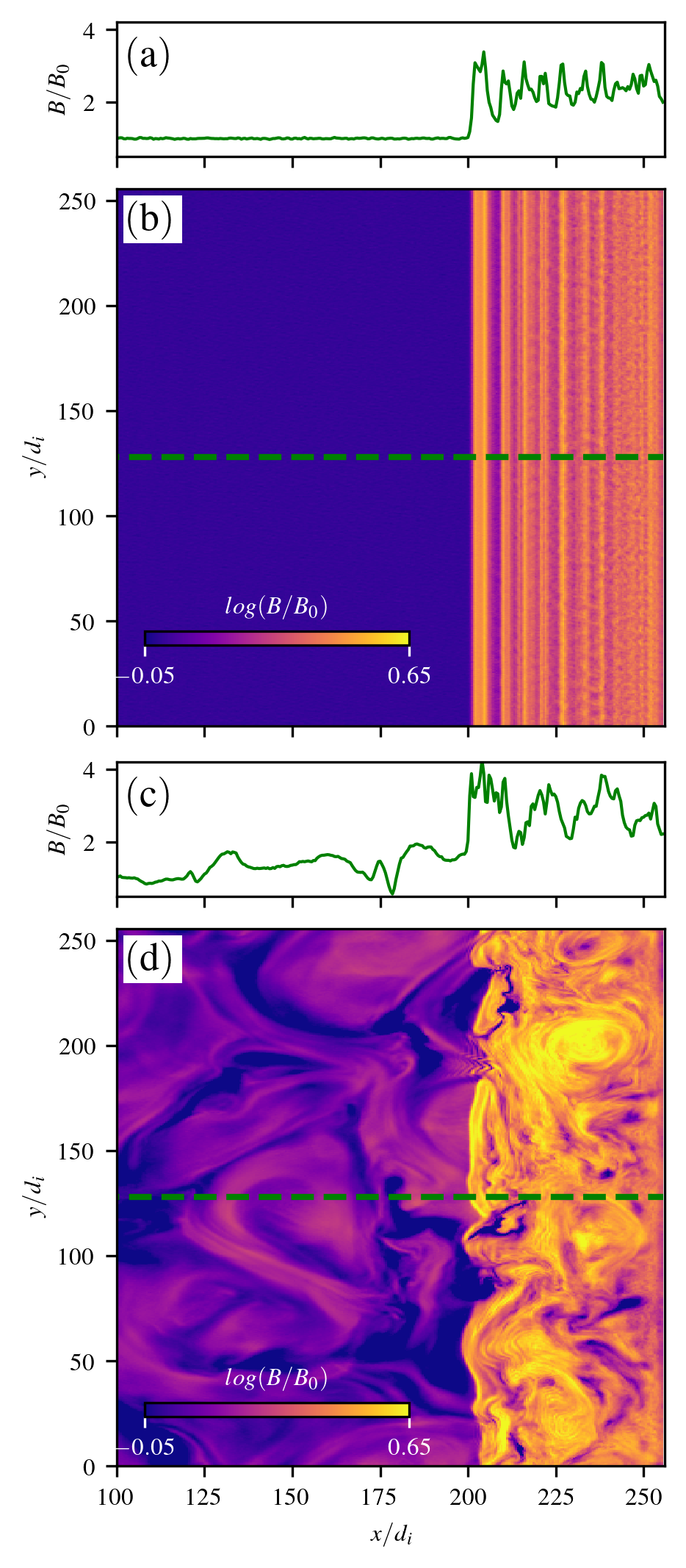}
\caption{Overview of unperturbed (top) vs perturbed (bottom) perpendicular shock simulations. Colormaps show simulation snapshots of magnetic field magnitude, and the one-dimensional $B/B_0$ plots in (a) and (c) correspond to horizontal cuts of the simulation domain along the green dashed lines in (b) and (d), respectively.\label{fig:f3_sim_ow}}
\end{figure}
 In the hybrid simulations, distances are normalised to the ion inertial length $d_i \equiv c/\omega_{pi}$, times to the inverse cyclotron frequency ${\Omega_{ci}}^{-1}$, velocity to the Alfv\'en speed $v_A$ (all referred to the unperturbed upstream state), and the magnetic field and density to their unperturbed upstream values, $B_0$ and $n_0$, respectively. The upstream magnetic field is in the out-of-plane direction, $\mathbf{B}_0 = B_0 \hat{z}$, consistent with the MHD simulation setup. For the upstream flow velocity, a value of  $V_\mathrm{in} = 2.5 v_A$  has been chosen, and the resulting Alfv\'enic Mach number of the shock is approximately $M_A \equiv v_{\rm{sh}}/v_A \simeq 3.3$. The upstream ion distribution function is an isotropic Maxwellian and the ion $\beta_i$ is 1.  The simulation $x-y$ domain  is (256 $\times$ 256) $d_i^2$, with spatial resolution $\Delta x$ = $\Delta y$ = 0.5 $d_i$. The time step for particle (ion) advance is $\Delta t$  = 0.01 $\Omega_{ci}^{-1}$. A small, nonzero resistivity is introduced in the magnetic induction equation, to limit excessive fluctuations at the grid-scale. The number of particles per cell used is always greater than 500 (upstream), keeping the  noise characteristic of PIC simulations, at a reasonable level.
\begin{figure*}[ht!]
\includegraphics[width=\textwidth]{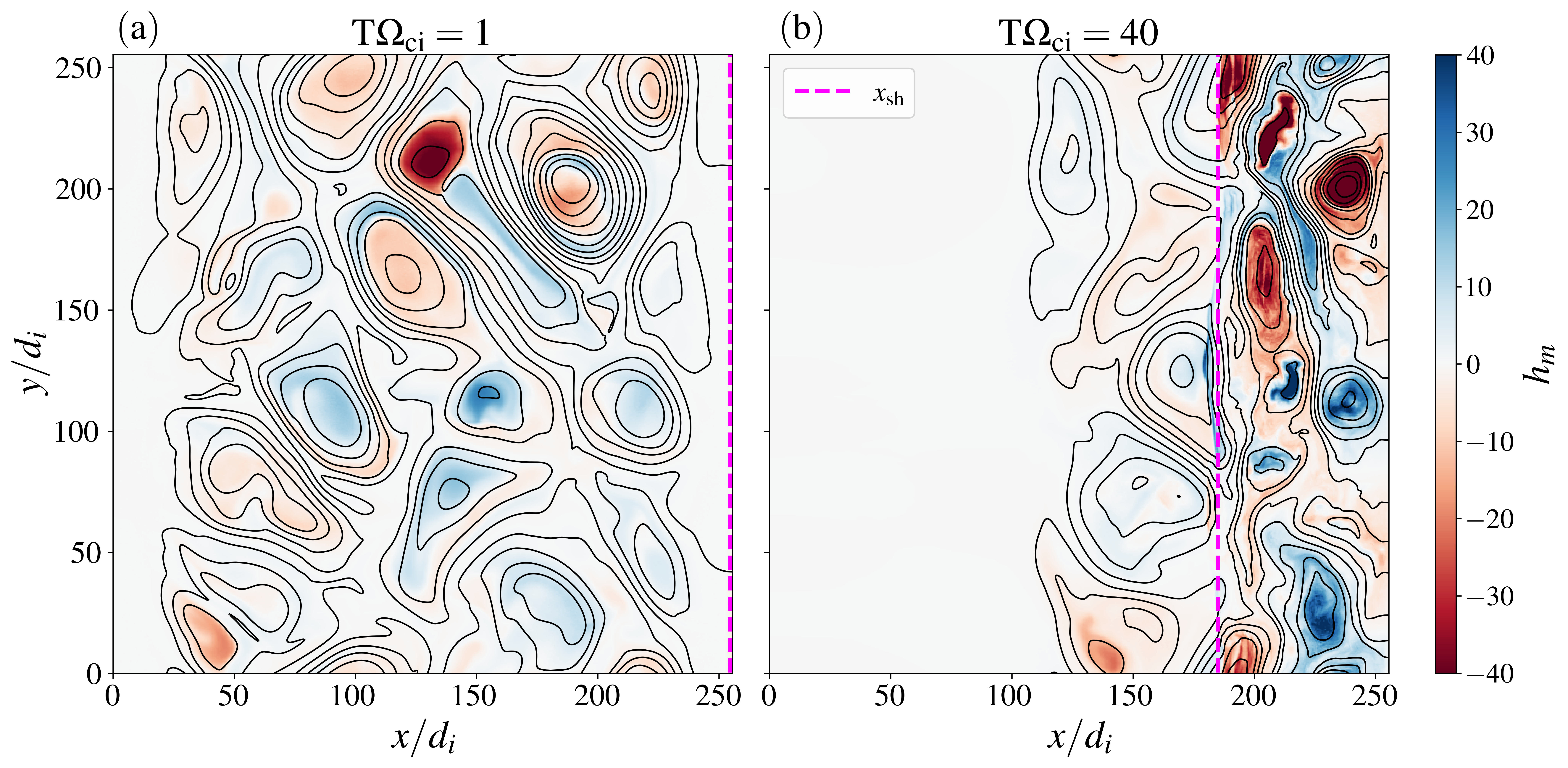}
\caption{Evolution of $h_m = a_z b_z$ for two different times in the simulation (color), with contours of vector potential $a_z$ (black curves). The vertical dotted magenta lines represent the nominal shock position. \label{fig:fig4_hm_sim}}
\end{figure*}

We present two different kinetic simulations, comparing an unperturbed case ($\delta b/B_0=0$) to a case including pre-existing turbulence in the shock upstream.
An overview of the simulations is shown in Figure \ref{fig:f3_sim_ow}. In the unperturbed case (panels (a) and (b)), the features of a low $M_A$, perpendicular shock are recovered, namely a ``quasi-laminar'' shock transition with quiet upstream conditions. The difference with respect to the case in which the shock propagates in a turbulent medium is striking (see Figure~\ref{fig:f3_sim_ow} (c)-(d)). Here, the shock front appears strongly perturbed by the incoming solar wind, inducing local departures from the nominal perpendicular geometry, an important ingredient for particle acceleration \citep[e.g.,][]{Johlander2016,Guo2015, Trotta2019}.  Furthermore, it is possible to observe how the incoming (upstream) turbulence is processed in the shock crossing, with structures being compressed along the shock normal direction and an increase of fluctuations in the shock downstream.

At this stage, it is interesting to address the behavior of turbulent helical structures through their transmission across the shock. Note that, because of the 2.5D geometry, the vector potential retains only the component perpendicular to the plane $(a_z)$; thus, magnetic helicity density can be evaluated as $h_m = a_z b_z$. Figure~\ref{fig:fig4_hm_sim} shows color maps of magnetic helicity with contours of the vector potential $a_z$ (black lines) at the initial simulation time and at $\rm T \Omega_{ci} = 40$. In the initial condition, corresponding to fully developed turbulence, helical structures can be easily identified (panel (a)). Important features involving turbulent structures can be identified by looking at the shock downstream in panel (b). In general, structures appear compressed after the shock crossing, and their helicity content appears enhanced with respect to the upstream, in agreement with the spacecraft observations. Immediately after the shock, it can be seen that the structures are elongated along the shock surface, due to the fact that the shock compression is along the shock normal ($\hat{x}$). It is interesting to note that most of the turbulent structures ``survive'' their passage across the shock, even though their dynamics is modified by the shock compression.

\begin{figure*}[ht!]
\includegraphics[width=\textwidth]{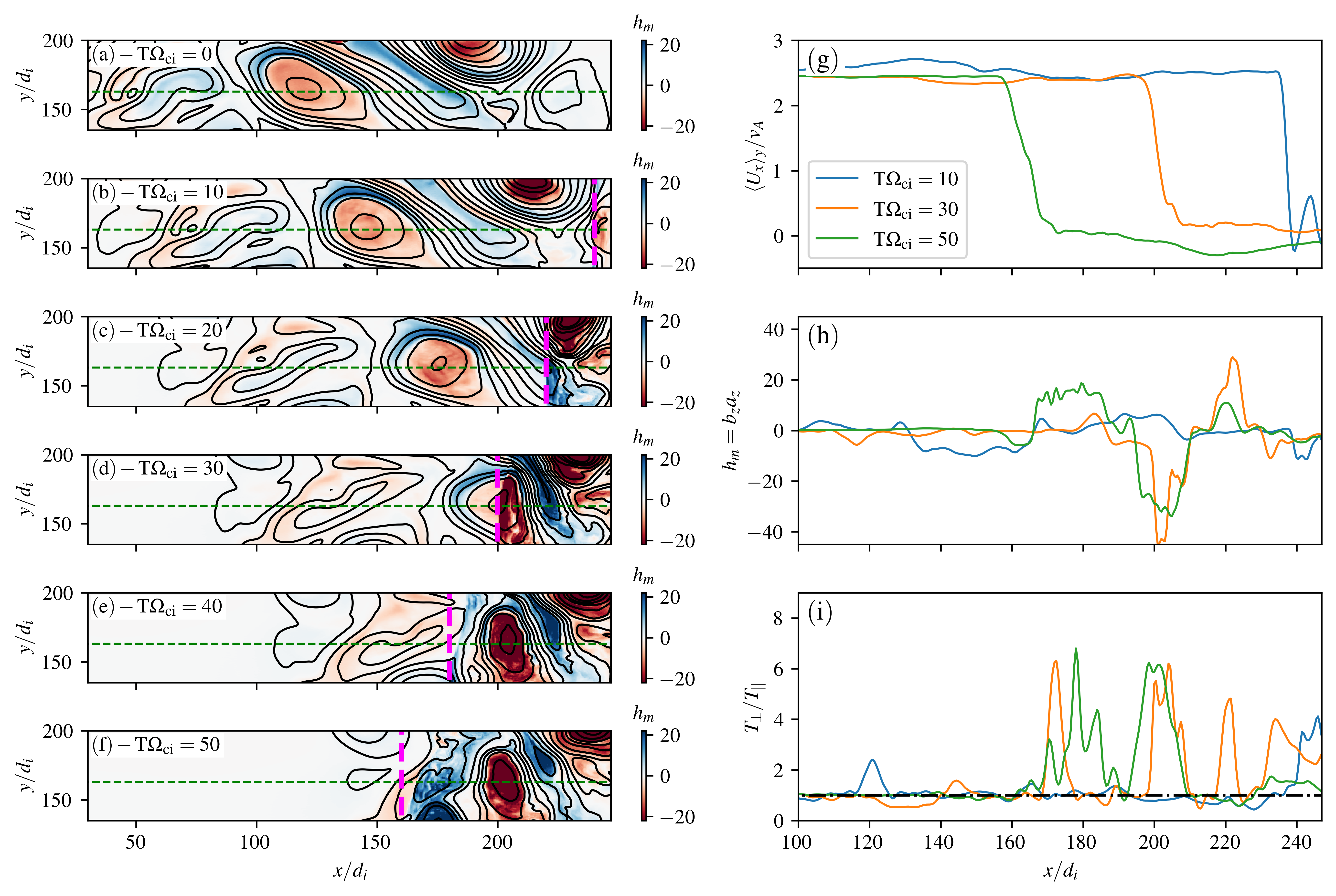}
\caption{(a)-(f): Helicity (color) and contours of the magnetic potential for a portion of the simulation domain, focusing on the evolution of a turbulent structure from upstream to downstream. The vertical (magenta) dashed lines identify the nominal shock position at that simulation time. (g)-(i): Horizontal cuts of the simulation domain along the $y = 160 \, d_i$ line (green dashed line in (a) - (f)), taken at three different times. The variables shown are the ion bulk flow speed normal to the shock and averaged along the $\hat{y}$ direction $\langle U_x\rangle_y$ (g), the magnetic helicity (h) and the perpendicular temperature anisotropy ((i), the dotted-dashed line represents $T_\perp/T_{||} =1$).  \label{fig:fig5_struct}}
\end{figure*}

So far, we focused on the overall behavior of turbulence transmission. In Figure~\ref{fig:fig5_struct} we introduce a ``Lagrangian'' approach, following the evolution of a turbulent structure from upstream to downstream. In the left panels, the passage of the structure initially located between 100 and 150 $d_i$ is highlighted, revealing the behavior described above, namely the structure survival through the shock passage, its compression (the structure size along the $\hat{x}$-direction reduces from roughly 50 $d_i$ down to around 20 $d_i$) and the subsequent increase of magnetic helicity. In the right panels of Figure~\ref{fig:fig5_struct}, we show one-dimensional cuts along the simulation domain, to provide a counterpart to the spacecraft observations discussed above. The panel (g), showing the ion bulk flow speed normal to the shock and averaged along the $\hat{y}$ direction $\langle U_x\rangle_y$, marks the (average) shock position at three different times. In panels (h) and (i), it is possible to observe the increase in helicity content for the structures, as well as the generation of a strong perpendicular temperature anisotropy. The helicity signal shown in panel (h), evaluated as $b_za_z$, is very similar --not shown here-- to the one obtained using the reduced measurement in Eq.~\ref{eq:Hm}, with increments taken along the virtual spacecraft trajectory (green-dashed lines of panels (a)--(f)).This behavior is in good agreement with the statistical behavior of observations, as discussed above.

\section{Discussion and conclusions}
\label{sec:conclusions}

We studied the transmission of turbulence across the perpendicular bow shock,
a crucial topic to understand how shocks interact with turbulence in the heliosphere and in other astrophysical settings \citep{Bykov2019,Guo2021}. We focus on the transmission of turbulent structures across the shock, with particular attention on their magnetic helicity, an important MHD rugged invariant \citep[][]{matthaeus1982evaluation}, which measures the local topology of the magnetic field.

An alignment between Wind (upstream) and MMS (crossing the quasi-perpendicular bow shock) has been used to sample the upstream and downstream conditions. Analogously, in simulations we observe, from upstream to downstream, an increase of the helicity content of turbulent structures, as well as an injection of perpendicular temperature anisotropy.

Two hybrid simulations have been performed, with parameters compatible with the ones observed \emph{in-situ}, comparing the behavior of a perpendicular, subcritical shock propagating in a laminar and turbulent medium. A striking difference between the two cases is found, with the unperturbed case exhibiting a laminar shock front and an ordered structure. When pre-existing turbulence is introduced in the upstream, the shock front is strongly distorted, with  the typical features of  quasi-perpendicular shock transitions not evident anymore. Another feature found both in observations and simulations is an increase of perpendicular temperature anisotropy in the ``shocked'' turbulent structures. We note that the behaviour observed here for the turbulent structures may play a fundamental role for particle acceleration -- a topic of growing interest in recent literature \citep[e.g.][]{Nakanotani2021, Nakanotani2022}.  The temperature anisotropy embedded into these structures is large in the simulation cases, a known effect of the simulations' reduced dimensionality \citep[see][for further details]{Burgess2007}. The problem will be investigated in full three-dimensional geometries in future works.

The simulations, in agreement with spacecraft observations, show that the transmitted turbulent structures are compressed during the shock penetration, with a subsequent increase in their magnetic helicity content.  It is possible to discuss this inhomogeneous helicity-injection mechanisms by taking in consideration the ideal, compressible MHD equations. Note that, for simplicity, here we take into account only the large scale contribution, neglecting the Hall term \citep{ServidioEA2007,PezziEA17}. From the induction equation $\frac{\partial {\bm b}}{\partial t}={\bm \nabla}\times\left[{\bm u}\times{\bm b}\right]$, writing down the analogous expression for the magnetic potential $\bm a$, assuming an ideal plasma in a 2.5D geometry, one gets

\begin{equation}
\left[\frac{\partial }{\partial t} + {\bm u}\cdot {\bm \nabla}\right] h_m = - h_m {\bm \nabla}\cdot{\bm u} + \left[{\bm \nabla} u_z \times {\bm \nabla}\left(\frac{a_z^2}{2}\right)\right]\cdot {\hat{\bm z}},
\label{eq:dHdt}
\end{equation}
where $\nabla\equiv(\frac {\partial}{\partial x},\frac {\partial}{\partial y}, 0 )$ is the in-plane gradients. From the above equation, note that one can integrate over a closed flux surface (bounded by a contour line of $a_z$) and obtain an expression for the total magnetic helicity. In the reference frame of the incoming structure, we measured (not shown here) that $\left\langle |h_m {\bm \nabla}\cdot{\bm u}| \right\rangle \gg \left\langle|\left[{\bm \nabla} u_z \times {\bm \nabla}\left(\frac{a_z^2}{2}\right)\right]\cdot {\hat{\bm z}}|\right \rangle$, confirming the helicity enhancement mechanism at the shock-front. This  amplification is clearly visible from the time-history of the structures when they encounter the discontinuity, in Figure \ref{fig:fig5_struct}. The relation between temperature anisotropy, magnetic helicity, and the compression and vorticity introduced by the shock transition may be a crucial aspect of the overall energetics of such systems and will be the object of future investigation.

\begin{acknowledgments}
Part of this work was performed using the DiRAC Data Intensive service at Leicester, operated by the University of Leicester IT Services, which forms part of the STFC DiRAC HPC Facility (www.dirac.ac.uk), under the project “dp031 Turbulence, Shocks and Dissipation in Space Plasmas”. The equipment was funded by BEIS capital funding via STFC capital grants ST/K000373/1 and ST/R002363/1 and STFC DiRAC Operations grant ST/R001014/1. DiRAC is part of the National e-Infrastructure. This work has received funding from the European Unions Horizon 2020 research and innovation programme under grant agreement No. 776262 (AIDA, www.aida-space.eu). This work has received funding from the European Unions Horizon 2020 research and innovation programme under grant agreement No. 101004159 (SERPENTINE, www.serpentine-h2020.eu).
This research is partially supported by the Parker Solar Probe mission through the IS$\odot$IS Theory and Modeling team and a subcontract from Princeton University (SUB0000317), NASA PSP Guest Investigator grants 80NSSC21K1765.
\end{acknowledgments}



\end{document}